%% file: belle-conf-0446.tex
\documentclass[aps,prl,preprint,tightenlines,superscriptaddress,showpacs,byrevtex]{revtex4}
\usepackage{epsfig, amssymb}

\begin{document}
\advance\hoffset by  -4mm

\newcommand{\de}{\Delta E}
\newcommand{\mbc}{M_{\rm bc}}
\newcommand{\bb}{B{\bar B}}
\newcommand{\qq}{q{\bar q}}
\newcommand{\ks}{\bar{K}^0}
\newcommand{\kshort}{K^0_S}
\newcommand{\kstar}{\bar{K}^{*0}}
\newcommand{\kz}{\bar{K}^{(*)0}}
\newcommand{\kpi}{K^-\pi^+}
\newcommand{\kpipin}{\kpi\pi^0}
\newcommand{\kpipipi}{\kpi\pi^-\pi^+}
\newcommand{\dkpi}{D^0\to\kpi}
\newcommand{\dkpipin}{D^0\to\kpipin}
\newcommand{\dkpipipi}{D^0\to\kpipipi}
\newcommand{\bdnks}{\bar{B}^0\to D^0 \ks}
\newcommand{\bdnkstar}{\bar{B}^0\to D^0 \kstar}
\newcommand{\bdsks}{\bar{B}^0\to D^{*0} \ks}
\newcommand{\bdskstar}{\bar{B}^0\to D^{*0} \kstar}
\newcommand{\bdnkz}{\bar{B}^0\to D^0 \kz}
\newcommand{\bdskz}{\bar{B}^0\to D^{*0} \kz}
\newcommand{\bdzks}{\bar{B}^0\to D^{(*)0} \ks}
\newcommand{\bdzkstar}{\bar{B}^0\to D^{(*)0} \kstar}
\newcommand{\bdzkz}{\bar{B}^0\to D^{(*)0} \kz}
\newcommand{\bdbarkstar}{\bar{B}^0\to\bar{D}^0 \kstar}
\newcommand{\bdsbarkstar}{\bar{B}^0\to\bar{D}^{*0} \kstar}
\newcommand{\bdzbarkstar}{\bar{B}^0\to\bar{D}^{(*)0} \kstar}
\newcommand{\dsdpi}{D^{*0}\to D^0\pi^0}
\newcommand{\bdppi}{\bar{B}^0\to D^+\pi^-}
\newcommand{\br}{{\cal B}}

\newcommand{\tim}{\times 10^{-5}}
\newcommand{\brbdnks}{3.72\pm 0.65\pm 0.37}
\newcommand{\brbdnkstar}{3.08\pm 0.56\pm 0.31}
\newcommand{\ulbdsks}{5.9}
\newcommand{\ulbdskstar}{4.8}
\newcommand{\brbdsks}{3.18^{+1.25}_{-1.12}\pm 0.32}
\newcommand{\brbdskstar}{2.34^{+1.24}_{-1.15}}
\newcommand{\brbdbarkstar}{0.51}
\newcommand{\brbdsbarkstar}{1.9}

\preprint{\vbox{ \hbox{   }
    \hbox{BELLE-CONF-0446}
    \hbox{ICHEP04 8-0693}
}}

\title{\Large \rm \quad\\[0.5cm]
Improved Measurements of $\bdnks$ and $\bdnkstar$ Branching
Fractions}
\input {author-conf2004.tex}
\noaffiliation

\begin{abstract}
We report on an improved study of $\bdzkz$ decays,
based on $274\times 10^6$ $\bb$ events
collected with the Belle detector at KEKB.
The following branching fractions have been measured:
$\br(\bdnks) =(\brbdnks)\tim$ and $\br(\bdnkstar) =(\brbdnkstar)\tim$.
We also obtain evidence with $3.2\sigma$ significance for $\bdsks$ 
with $\br(\bdsks)=(3.2^{+1.2}_{-1.1}\pm 0.4)\tim$.
No significant signal has been found for the $\bdskstar$ and $\bdzbarkstar$ 
decay modes, and upper limits at 90\% CL are presented.
\end{abstract}
\maketitle

{\renewcommand{\thefootnote}{\fnsymbol{footnote}}}
\setcounter{footnote}{0}

Within the Standard Model, $CP$ violation arises due to a
single phase in the Cabibbo-Kobayashi-Maskawa quark mixing
matrix~\cite{km}. Measurements of the Unitary Triangle angles therefore
test the consistency of the Standard Model. 
Precise measurements of the branching fractions for $\bdnkstar$, 
$\bdbarkstar$ and ${\bar B^0}\to D^0_{CP} {\bar{K}^{*0}}$ decays, 
where $D^0_{CP}$ denotes $D^0$ or $\bar{D}^0$ decay to a CP eigenstate, 
will allow a measurement of the angle $\phi_3$~\cite{dunietz}.
The decay $\bdnks$ can also be used to measure time-dependent
CP asymmetry in $B$ decays~\cite{gronau}.
The $\bdzks$ and $\bdzkstar$ decays have been previously observed by
Belle~\cite{belle_br}.
For all these measurements, knowledge of the ratio of amplitudes 
$R_{D^0K^{*0}}=\frac{A(\bdbarkstar)}
{A(\bdnkstar)}=\sqrt{\frac{\br(\bdbarkstar)}{\br(\bdnkstar)}}$
is of crucial importance;
its value is predicted to be about 0.4~\cite{fleischer}.

Here we report improved measurements of $\bdzks$ and
search for $\bdzkstar$ and $\bar{B}^0\to \bar{D}^{(*)0}\kstar$~\cite{conj}
decays with the Belle detector~\cite{NIM} at the KEKB asymmetric energy 
$e^+e^-$ collider~\cite{KEKB}. The results are based on 
a data sample collected at the center-of-mass (CM) energy of the 
$\Upsilon(4S)$ resonance, which  contains $274\times 10^6$ produced 
$\bb$ pairs. 

The Belle detector is a large-solid-angle magnetic
spectrometer that consists of a silicon vertex detector (SVD),
a 50-layer central drift chamber (CDC), an array of
aerogel threshold \v{C}erenkov counters (ACC),
a barrel-like arrangement of time-of-flight
scintillation counters (TOF), and an electromagnetic calorimeter (ECL)
comprised of CsI(Tl) crystals located inside
a superconducting solenoid coil that provides a 1.5~T
magnetic field.  An iron flux-return located outside of
the coil is instrumented to detect $K_L^0$ mesons and to identify
muons (KLM).  The detector is described in detail elsewhere~\cite{NIM}.
Two different inner detector configurations were used. For the first sample
of 152 million $B\bar{B}$ pairs, a 2.0 cm radius beampipe
and a 3-layer silicon vertex detector were used;
for the latter 122 million $B\bar{B}$ pairs,
a 1.5 cm radius beampipe, a 4-layer silicon detector
and a small-cell inner drift chamber were used\cite{Ushiroda}.

Charged tracks are selected with a set of requirements based on the
average hit residual and impact parameter relative to the
interaction point (IP). We also require that the transverse momentum of
the tracks be greater than 0.1 GeV$/c$ in order to reduce the low 
momentum combinatorial background.

For charged particle identification (PID), the combined information
from specific ionization in the central drift chamber ($dE/dx$),
time-of-flight scintillation counters (TOF) and aerogel \v{C}erenkov
counters (ACC) is used.
At large momenta ($>2.5$~GeV$/c$) only the ACC and $dE/dx$ are used.
Charged kaons are selected with PID criteria that have
an efficiency of 88\%, a pion misidentification probability of 8\%,
and negligible contamination from protons.
All charged tracks having PID consistent with the pion
hypothesis that are not identified as electrons are considered
as pion candidates.

Neutral kaons are reconstructed via the decay $K_S^0\to\pi^+\pi^-$
with no PID requirements for these pions.
The two-pion invariant mass is required to be within 6~MeV$/c^2$
($\sim 2.5\sigma$) of the nominal $K^0$ mass and the displacement of
the $\pi^+\pi^-$ vertex from the IP in the transverse
($r-\phi$) plane is required to be between 0.2~cm and 20~cm. 
The direction from the IP to the $\pi^+\pi^-$ vertex is required to 
agree within 0.2 radians in the $r-\phi$ plane with the combined 
momentum of the two pions.

A pair of calorimeter showers not associated with charged tracks,
with an invariant mass within 15~MeV$/c^2$ ($\sim 3\sigma$) of the 
nominal $\pi^0$ mass is considered as a $\pi^0$ candidate. An energy 
deposition of at least 30~MeV and a photon-like shape are required 
for each shower.

$\kstar$ candidates are reconstructed from $\kpi$ pairs with an 
invariant mass within 50~MeV$/c^2$ ($1\Gamma$) of the nominal $\kstar$ mass.
The polarization of $\kstar$ mesons in $B$ decays is also utilized to
reject background through the use of the helicity angle $\theta_{K^*}$, 
defined as the angle between the $\kstar$ momentum in the $B$ meson
rest frame and the $K^-$ momentum in the $\kstar$ rest frame.
We require $|\cos(\theta_{K^*})|>0.3$ for $\bdzkstar$
decay channel. This rejects about 20\% of the background and 
retains 97\% of the signal. For the $\bdzbarkstar$ the expected 
signal to background ratio is worse and we require 
$|\cos(\theta_{K^*})|>0.5$.

We reconstruct $D^0$ mesons in the decay channels:
$\kpi$, $\kpipipi$ and $\kpipin$, using requirements that the
invariant mass be within 20~MeV$/c^2$, 15~MeV$/c^2$ and 25~MeV$/c^2$
($\sim 3\sigma$) of the  nominal $D^0$ mass, respectively.
In each channel we further define a $D^0$ mass sideband region,
with a width twice that of the signal region and location within 
0.1~GeV$/c^2$ on the nominal $D^0$ mass.
For the $\pi^0$ from the $\dkpipin$ decay, we require that 
its momentum in the CM frame be greater than 0.4~GeV$/c$ 
in order to reduce combinatorial background.
$D^{*0}$ mesons are reconstructed in the $\dsdpi$ decay mode. 
The mass difference between $D^{*0}$ and $D^0$ candidates is required 
to be within 4~MeV$/c^2$ of the expected value ($\sim 4\sigma$).

We combine $D^{(*)0}$ candidates with $K^0_S$ or $\kstar$ candidates
to form $B$ 
mesons. Signal events are identified by their CM energy difference, 
$\de=(\sum_iE_i)-E_{\rm b}$, and the beam constrained mass, 
$\mbc=\sqrt{E_{\rm b}^2-(\sum_i\vec{p}_i)^2}$, where $E_{\rm b}$ is
the beam energy and $\vec{p}_i$ and $E_i$ are the momenta and energies 
of the $B$ meson decay products in the CM frame.
We select events with $\mbc>5.2$~GeV$/c^2$ and $|\de|<0.2$~GeV,
and define a $B$ signal region of
$5.272$~GeV$/c^2<\mbc<5.288$~GeV$/c^2$ and $|\de|<0.03$~GeV.
In the rare cases where there is more than one  candidate in an 
event, the candidate with the $D^{(*)0}$ and $\kz$ masses 
closest to their nominal values is chosen.
We use Monte Carlo (MC) simulation to model the response of
the detector and determine the efficiency~\cite{GEANT}.

To suppress the large combinatorial background dominated by 
the two-jet-like $e^+e^-\to\qq$ continuum 
process, variables that characterize the event topology are used. 
We require $|\cos\theta_{\rm thr}|<0.80$, where $\theta_{\rm thr}$ is 
the angle between the thrust axis of the $B$ candidate and that of the 
rest of the event. This requirement eliminates 77\% of the continuum 
background and retains 78\% of the signal. We also construct a 
Fisher discriminant, ${\cal F}$, which is based on the production
angle of the $B$ candidate, the angle of the $B$ candidate thrust axis 
with respect to the beam axis, and nine parameters that characterize 
the momentum flow in the event relative to the $B$ candidate thrust 
axis in the CM frame~\cite{VCal}. We impose a requirement on 
${\cal{F}}$ that rejects 67\% of the remaining continuum background 
and retains 83\% of the signal. 
For the suppressed $\bdzbarkstar$ modes we make a
stricter requirement, which has 45\% efficiency and 93\% background 
rejection.

Among other $B$ decays, the most serious background comes from 
$B^0\to D^-\pi^+$, $D^-\to \kz K^-$, $\kz K^-\pi^0$,
$\kz K^-\pi^-\pi^+$ and $B^0\to D^-K^+$, $D^-\to \kz\pi^-$, 
$\kz\pi^-\pi^0$, $\kz\pi^-\pi^-\pi^+$. 
These decays produce the same final state as the $\bdzkz$ signal,
and their product branching fractions are up to ten times higher than 
those expected for the signal.
To suppress this type of background, we exclude candidates if the 
invariant mass of the combinations listed above is consistent
with the $D^-$ hypothesis within 25~MeV/$c^2$ ($\sim 3\sigma$).
The $\bar{B}^0\to D^{*+}K^-$, $D^{*+}\to D^0\pi^+$ decay can also
produce the same final state as the $\bdnkstar$ decay; however this decay 
is kinematically separated from the signal - the invariant mass selection
criteria for $\kstar$ candidates completely eliminates this background.
Another potential $\bb$ background comes from the
$\bar{B}^0\to D^{(*)0}\rho^0$ decay channel with one pion from the 
$\rho^0$ decay
misidentified as a kaon. The reconstructed $\kpi$ invariant mass 
spectra for these events overlap with the $\kstar$ signal mass region, 
while their $\de$ distribution is shifted by about 70~MeV$/c^2$.
We study this background using MC simulation. The contribution to the 
$\bdzkstar$ signal region is found to be less than 0.2 events.
We examined the possibility that other $B$ meson decay modes might
produce backgrounds that peak in the signal region by studying a large
MC sample of generic $\bb$ events.
No other peaking backgrounds were found.

\begin{figure*}
  \includegraphics[width=0.45\textwidth] {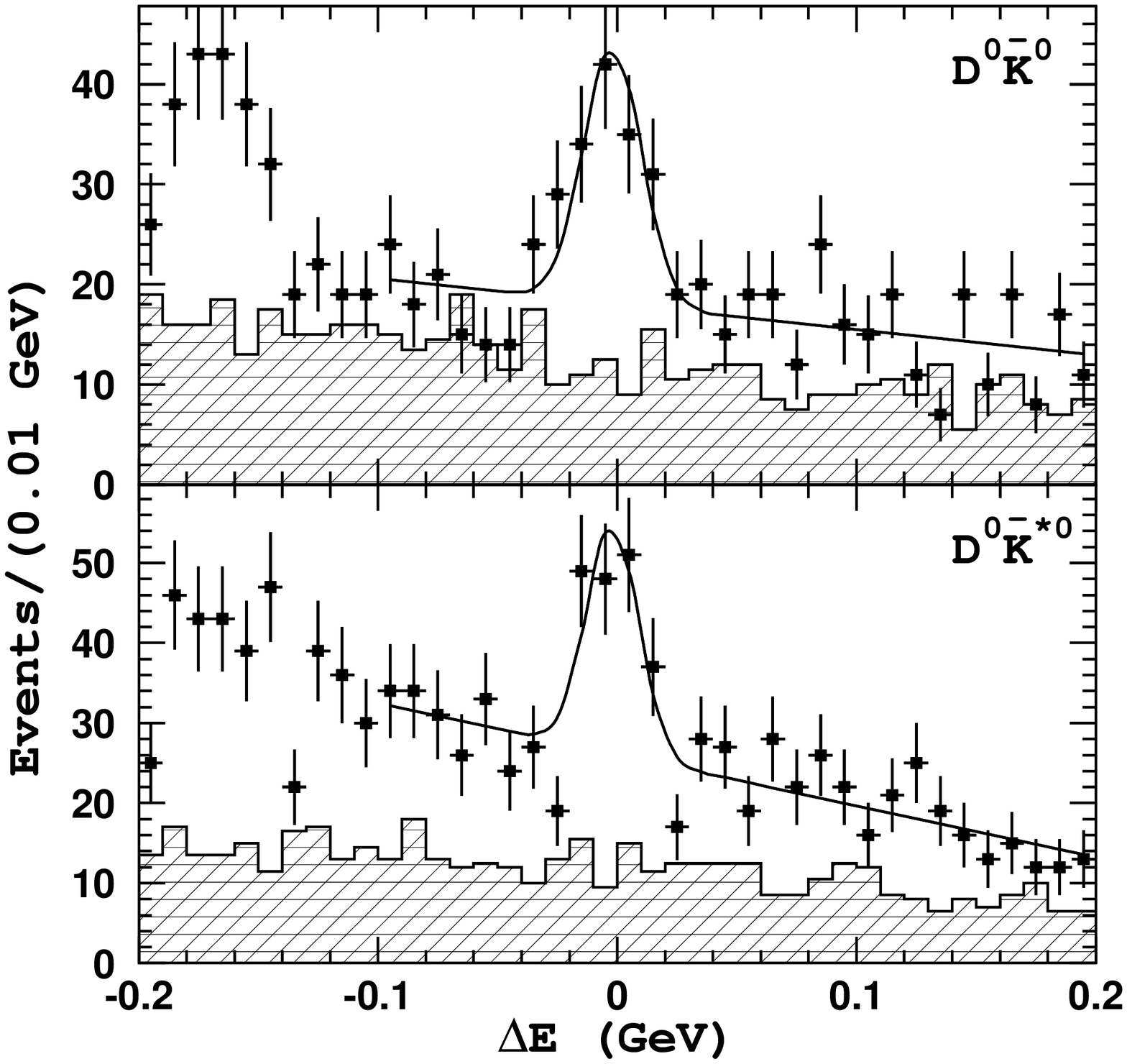} \hfill
  \includegraphics[width=0.45\textwidth] {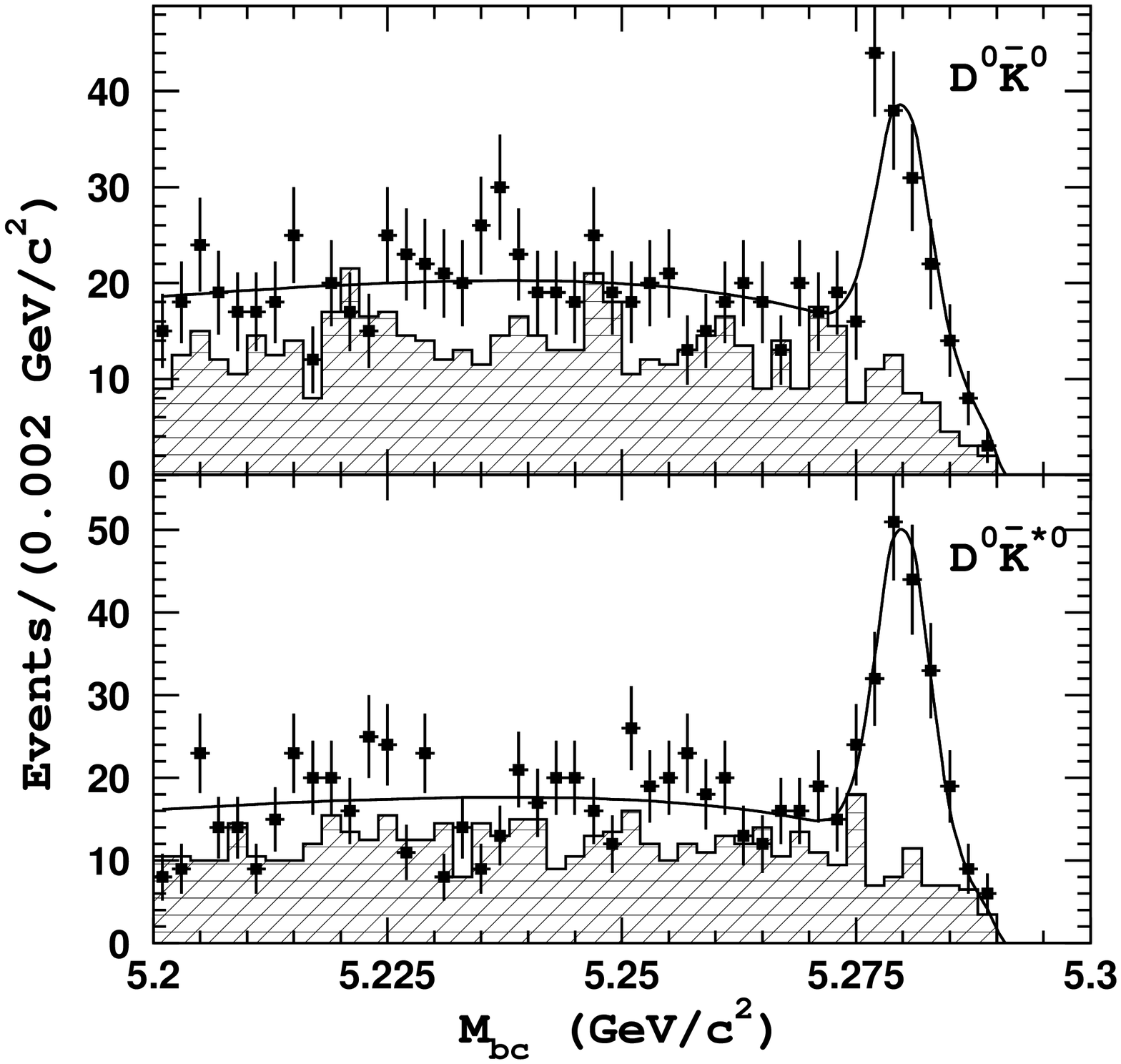} 
  \caption{$\de$ (left) and $\mbc$ (right) distributions for the 
    $\bdnkz$ candidates. Points with errors
    represent the experimental data, hatched histograms show
    the $D^0$ mass sidebands and curves are the results of the fits.}
  \label{mbc_all}
\end{figure*}

\begin{figure*}
  \includegraphics[width=0.24\textwidth] {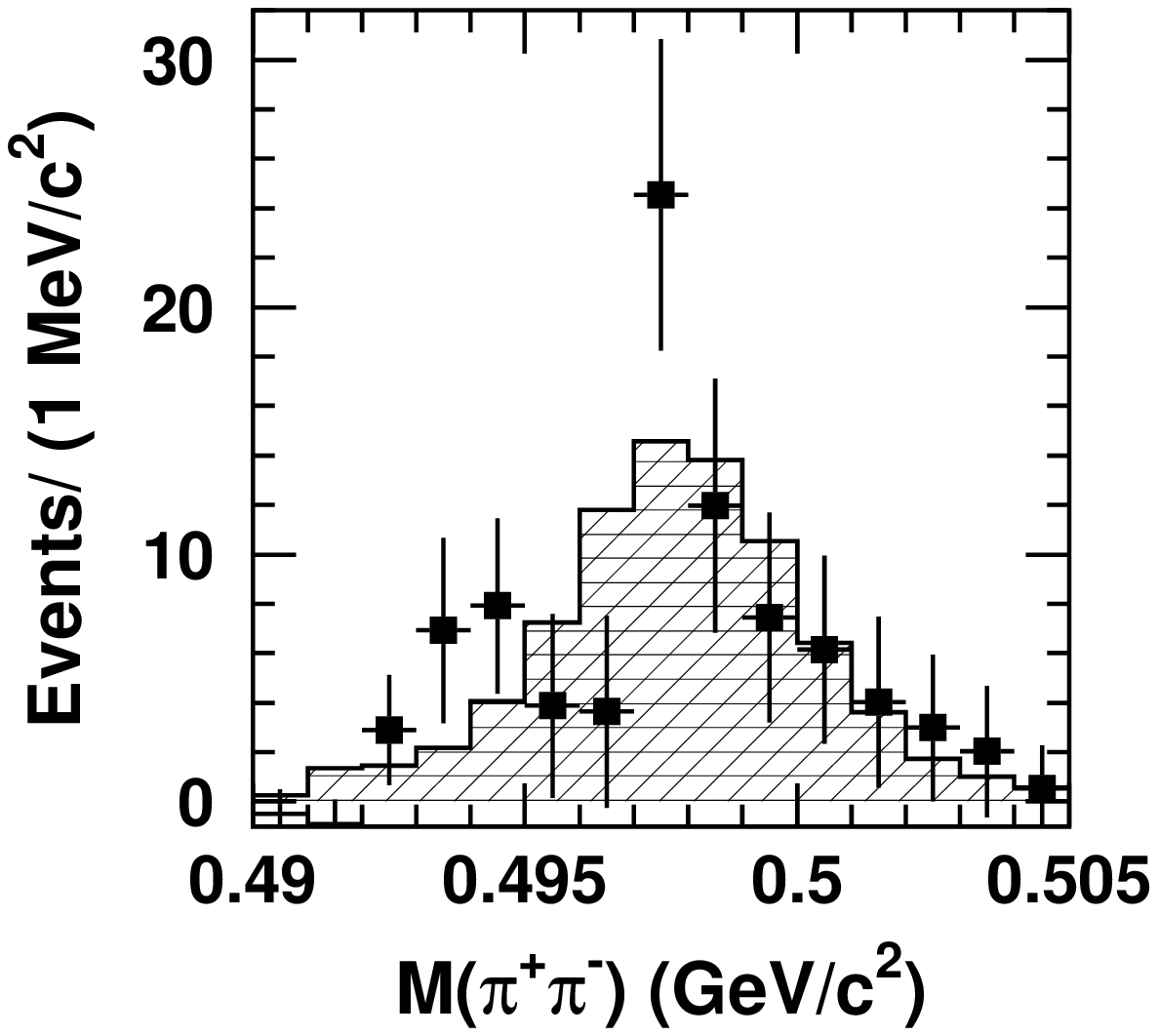} 
  \includegraphics[width=0.24\textwidth] {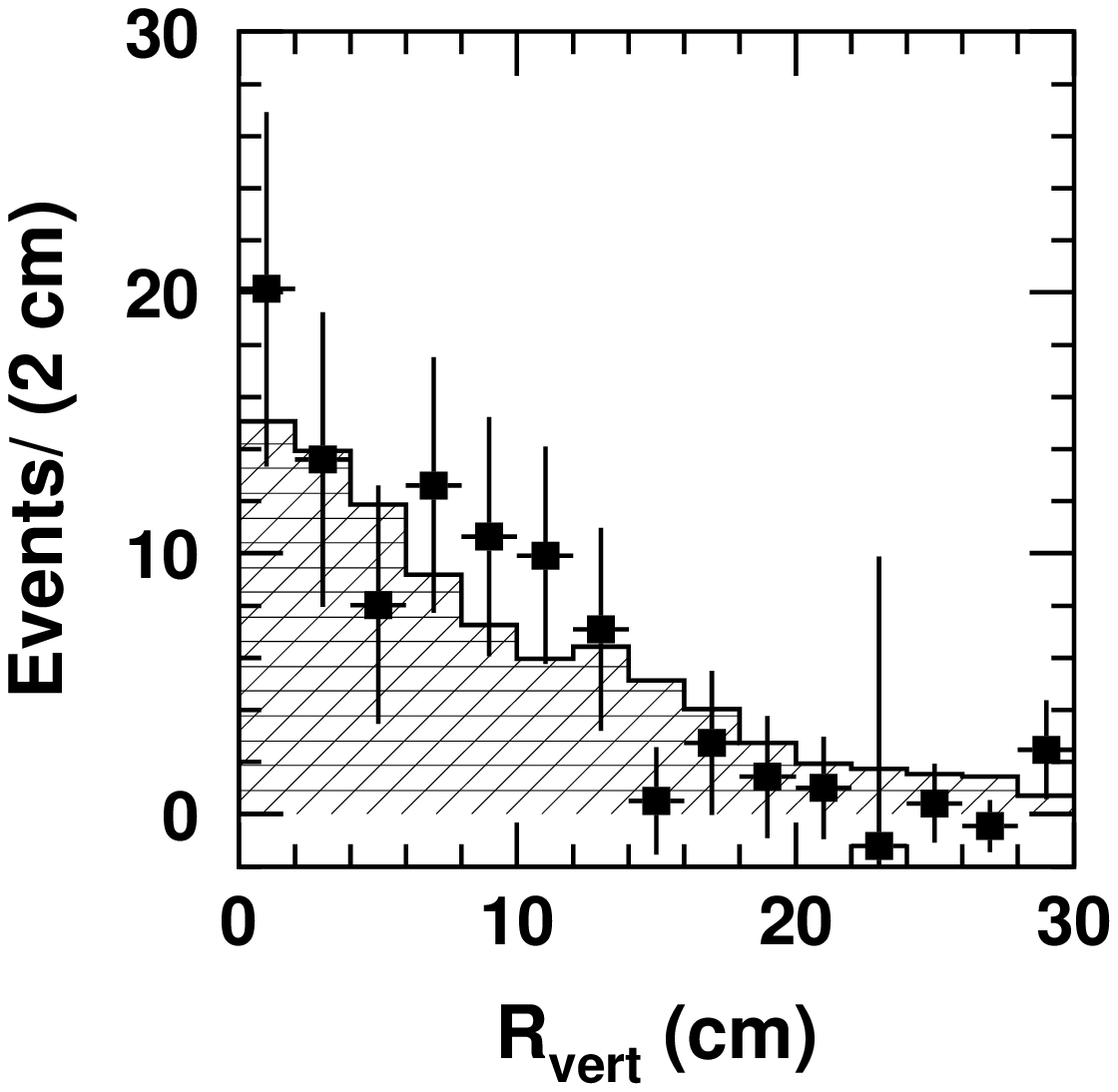}
  \includegraphics[width=0.24\textwidth] {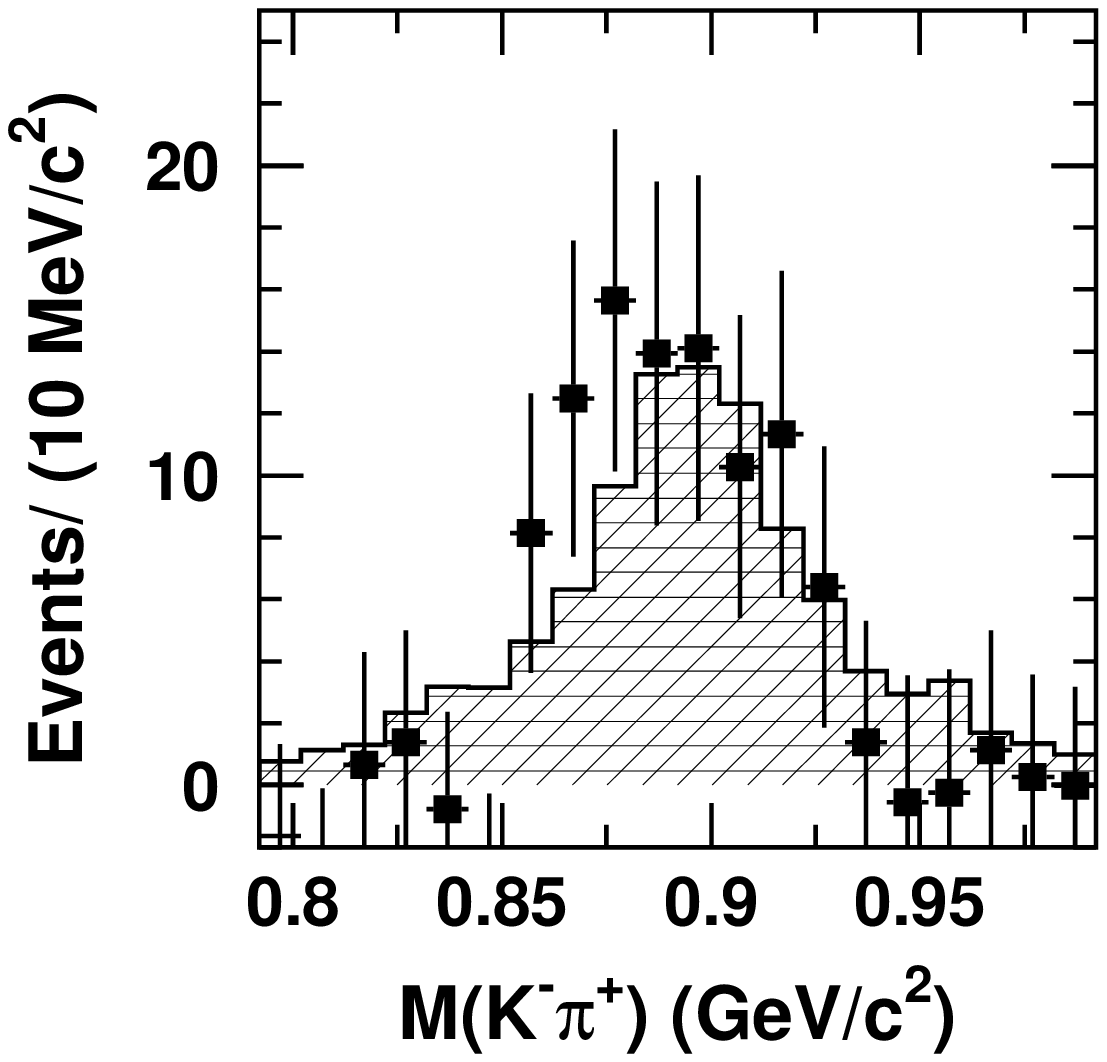} 
  \includegraphics[width=0.24\textwidth] {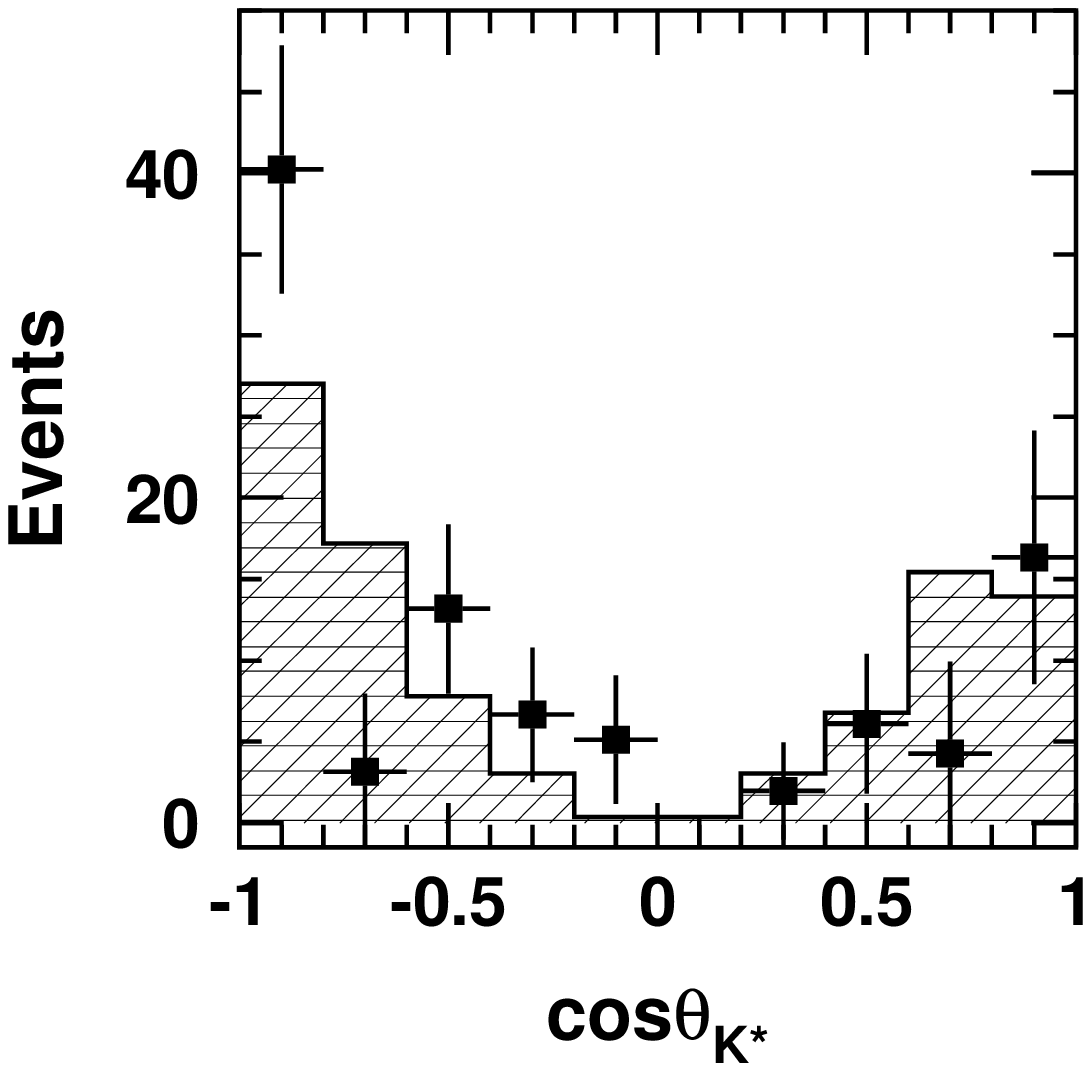}
  \caption{From left to right: Distributions of  invariant mass and
    flight distance for the $K^0_S$ candidates' in $\bdnks$ channel, 
    invariant mass and helicity distributions for the  $\kstar$
    candidates' in $\bdnkstar$ channel.
}
  \label{check}
\end{figure*}

The $\de$ and $\mbc$ distributions for $\bdnkz$ candidates are
presented in Fig.~\ref{mbc_all}, where all three $D^0$ decay modes are 
combined. 
Each distribution is made for events from the signal
region of the other parameter.
The hatched histograms in Fig.~\ref{mbc_all} are
the distributions for events in the $D^0$ mass sideband
scaled according to the $D^0$ mass selection.
The sideband shape replicates the background shape well, confirming
that the background is mainly combinatorial in nature. 
Clear signals are observed for the $D^0\ks$ and $D^0\kstar$ final states.
As a cross check, we also study the $K^0_S$ candidates' 
invariant mass and flight distance distributions and $\kstar$ 
candidates' invariant mass and helicity distributions for these decays.
The distributions mentioned above are shown in Fig.~\ref{check}, where the
points with error bars are the results of fits to the $\de$ spectra for 
data in the corresponding bin, and the histograms are distributions
from signal MC. All distributions are consistent with the MC expectations.

For each $D^0$ decay mode, the $\de$ distribution is fitted with a 
Gaussian for signal and a linear function for background. The Gaussian 
mean value and width are fixed to the values from MC simulations of
the signal. The region $\de<-0.1$~GeV is excluded from the fit 
to avoid contributions from other $B$ decays, such as 
$B\to D^{(*)0}\kz (\pi)$ 
where $(\pi)$ denotes a possible additional pion.
For the $\mbc$ distribution fit we use the sum of a signal Gaussian
and an empirical background function with a kinematic 
threshold~\cite{argus}, with a shape parameter fixed from the analysis 
of the off-resonance data.
For the calculation of branching fractions, we use the signal yields
determined from fits to the $\de$ distribution. This minimizes the
possible bias from other $B$ meson decays, which tend to peak in $\mbc$
but not in $\de$.
The fit results are presented in Table~\ref{defit}, where the listed
efficiencies include intermediate branching fractions.  
The statistical significance of the signal quoted in Table~\ref{defit} 
is defined as
$\sqrt{-2\ln({\cal L}_0/{\cal L}_{\rm max})}$, where 
${\cal L}_{\rm max}$ and
${\cal L}_0$ denote the maximum likelihood with the nominal signal
yield and the signal yield fixed at zero, respectively.

\begin{figure*}
  \includegraphics[width=0.24\textwidth] {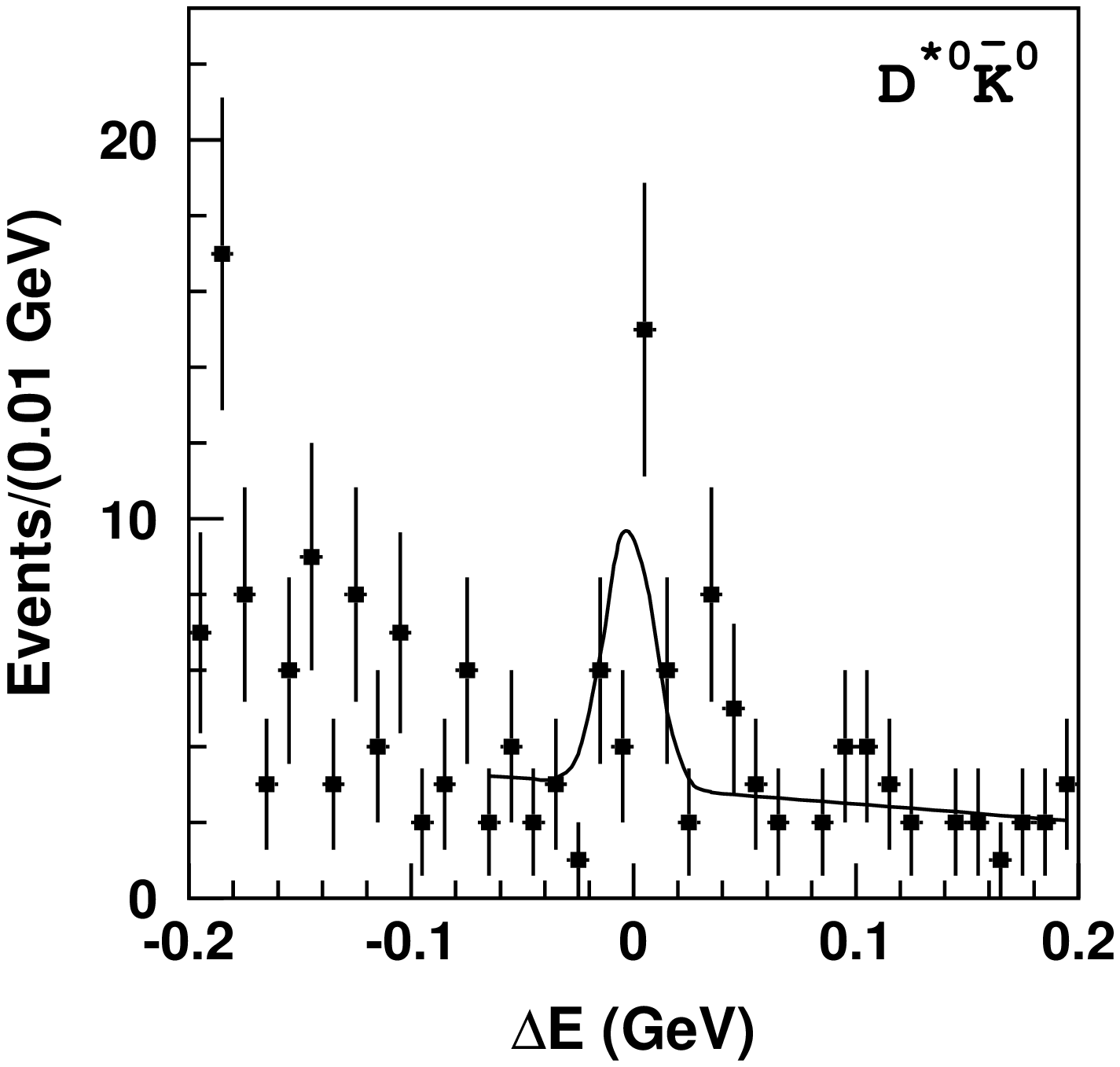}
  \includegraphics[width=0.24\textwidth] {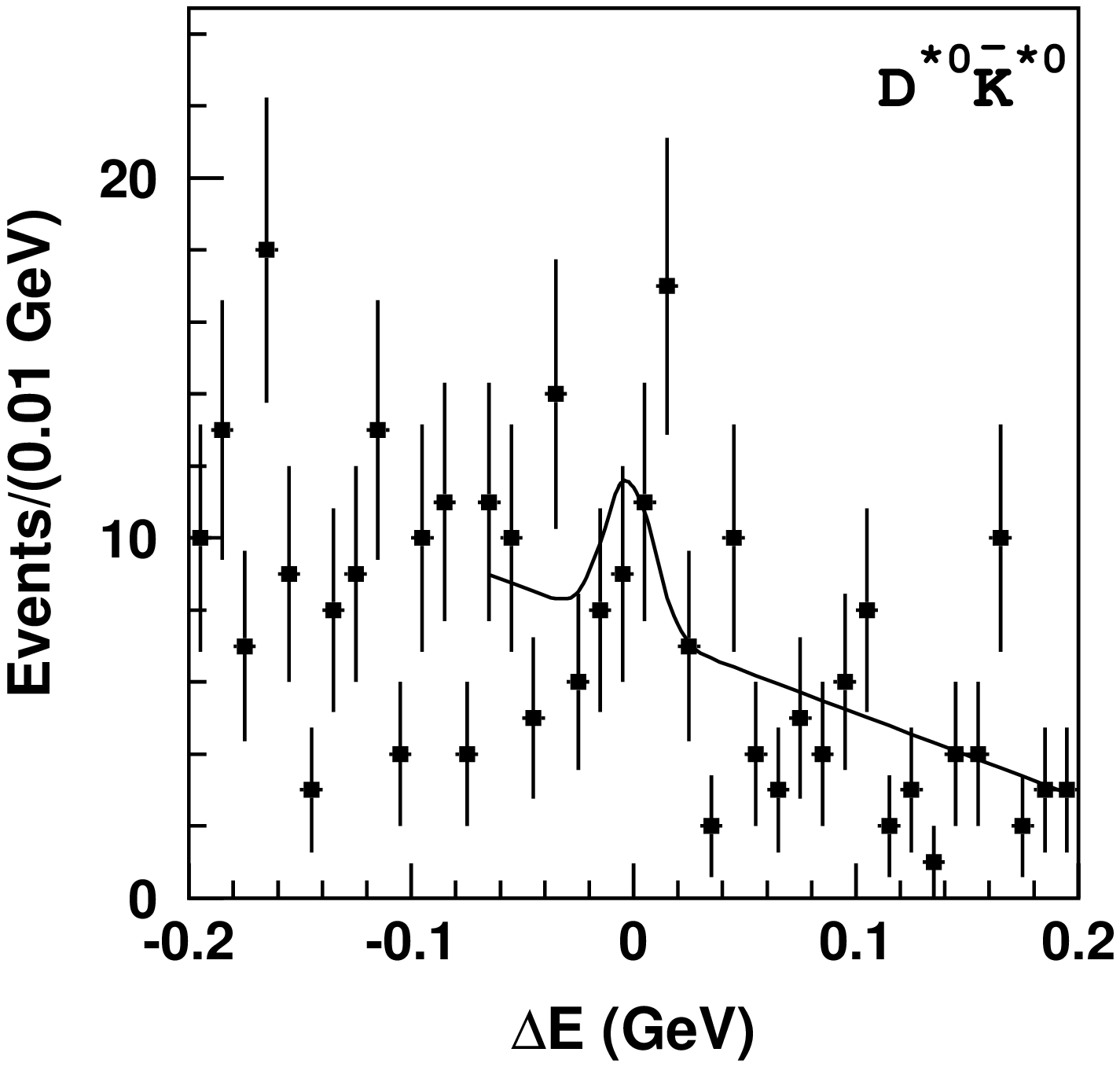}
  \includegraphics[width=0.24\textwidth] {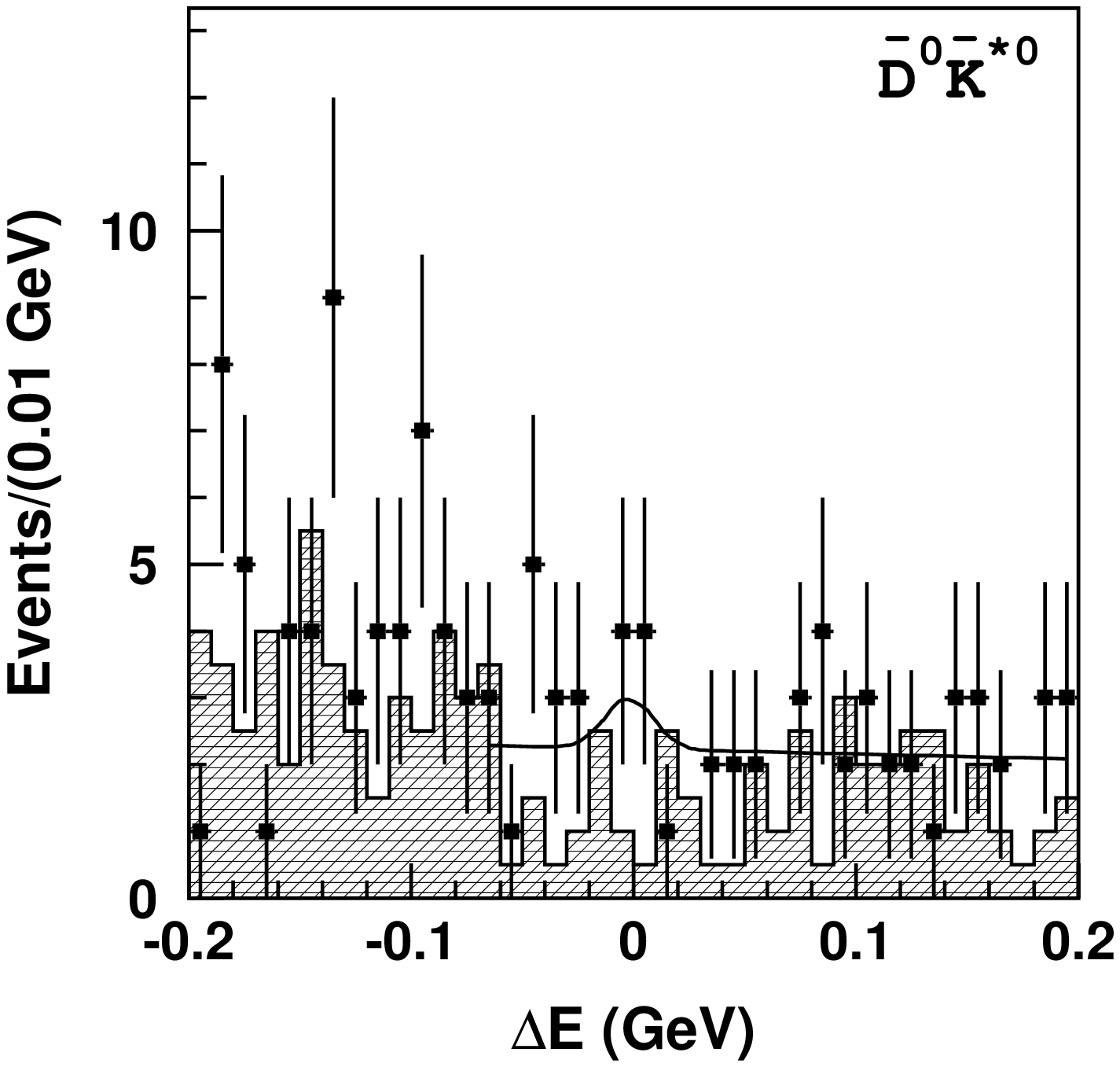}
  \includegraphics[width=0.24\textwidth] {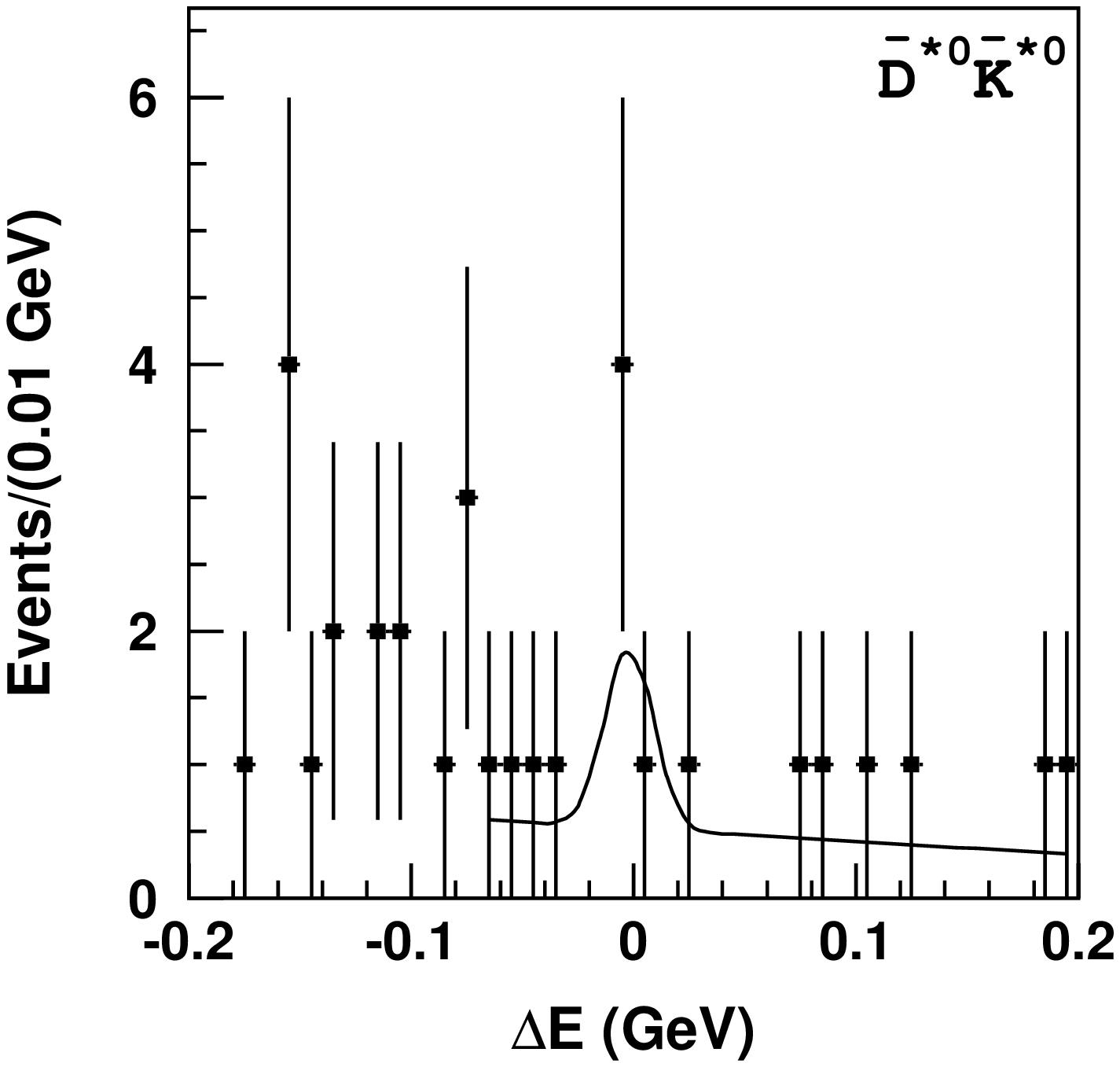}
  \caption{$\de$ distributions for the $\bdskz$ and $\bdzbarkstar$ 
    candidates. Points with error bars represent the experimental
    data, hatched histograms show the $D^0$ mass sidebands
    and curves show the results of the fits.}
  \label{de_ul}
\end{figure*}

\begin{table*}
\caption{Fit results, efficiencies, branching fractions and
    statistical significances for $\bdzkz$ decays.}
\footnotesize
\medskip
\label{defit}
  \begin{tabular*}{\textwidth}{l@{\extracolsep{\fill}}cccc}\hline\hline
 Mode  & $\de$ yield & 
Efficiency ($10^{-3}$) & $\br$ ($10^{-5}$) & Significance\\\hline

$\bdnks$ & $78.1\pm 14.1$ & 
$8.02$ & $\brbdnks$ & $6.6\sigma$ \\

$\bdnkstar$ & $77.7\pm 14.6$ & 
$9.98$ & $\brbdnkstar$ & $6.3\sigma$\\\hline

$\bdsks$ & $19.2^{+6.4}_{-5.8}$ & 
$1.97$ & $\brbdsks$ & $3.2\sigma$ \\

$\bdskstar$ & $12.3^{+7.5}_{-7.0}$ & 
$2.54$ & $\brbdskstar(<\ulbdskstar)$ 90\% CL & $2.1\sigma$ \\\hline

$\bdbarkstar$ & $0.4^{+3.6}_{-3.1}$ & 
$6.52$ & $<\brbdbarkstar$ 90\% CL & --- \\

$\bdsbarkstar$ & $3.3^{+2.7}_{-2.1}$ & 
$1.72$ & $<\brbdsbarkstar$ 90\% CL & --- \\\hline\hline

\end{tabular*}
\end{table*}

For the final result we use a simultaneous fit to the $\de$ distributions 
for the three $D^0$ decay channels taking into account the corresponding 
detection efficiencies.
The normalization of the background in each $D^0$ sub-mode is
allowed to float while the signal yields are required to satisfy the 
constraint
$N_i= N_{\bb}\cdot\br(\bdzkz)\cdot\varepsilon_i\, ,$
where the branching fraction $\br(\bdzkz)$ is a fit parameter;
$N_{\bb}$ is the number of $\bb$ pairs
and $\varepsilon_i$ is the efficiency, which includes all
intermediate branching fractions.

The statistical significance for $\bdsks$ is $3.2\sigma$. This is
the first evidence for this decay.
We do not observe significant signals for the $\bdskstar$ and the
$\bdzbarkstar$ decay modes and set 90\% CL upper limits for them. 
Figure~\ref{de_ul} shows the $\de$ distributions
for $\bdskz$ and $\bdzbarkstar$ candidates.
The upper limit $N$ is calculated from the relation 
$\int^N_0 {\cal L}(n) dn=0.9\int^{\infty}_0 {\cal L}(n) dn$, 
where ${\cal L}(n)$ is the maximum likelihood with the signal 
yield equal to $n$. We take into account the systematic uncertainties
in these calculations by reducing the detection efficiency by one
standard deviation. 
Using the measured branching fraction for $\bdnkstar$ decay and upper 
limit for $\bdbarkstar$, we set an upper limit on the amplitude ratio:
$R_{D^0K^{*0}}<0.39$.

As a check, we apply a similar procedure to the decay chains with the 
same final states: $\bar{B}^0\to D^+[K_S^0 K^+]\pi^-$ and 
$\bar{B}^0\to D^{*+}[D^0\pi^+]K^-$.
The estimated branching fractions
are consistent with the world average values~\cite{PDG}.

The following sources of systematic errors are found to be significant:
tracking efficiency (1-2\% per track), kaon identification
efficiency (1\%), $\pi^0$ efficiency (6\%), $K^0_S$ reconstruction 
efficiency (6\%), efficiency for slow pions 
from $\dsdpi$ decays (8\%), $D^{(*)0}$ branching fraction
uncertainties (2\% - 6\%), signal and background shape
parameterization (4\%) and MC statistics (2\% - 3\%).
The tracking efficiency error is estimated using
$\eta$ decays to $\gamma\gamma$ and $\pi^+\pi^-\pi^0$.
The kaon identification uncertainty is determined 
from $D^{*+}\to D^0\pi^+$, $D^0\to K^-\pi^+$ decays.
The $\pi^0$ reconstruction uncertainty is obtained using 
$D^0$ decays to $\kpi$ and $\kpipin$.
We assume equal production rates for $B^+B^-$ and $B^0\bar B^0$ pairs 
and do not include the uncertainty related to this assumption in the 
total systematic error. The overall systematic uncertainty is
found to be 10\% for $\bdnkz$ and 13\% for $\bdskz$.

In summary, we report improved measurements of $\bdnkz$ branching
fractions $\br(\bdnks)=(\brbdnks)\tim$ and 
$\br(\bdnkstar)=(\brbdnkstar)\tim$.
Note that we ignore the possible contribution of $\bar{B}^0\to D^0K^0$
to the former result, since we do not distinguish between 
$\ks$ and $K^0$. We also report the first evidence, with $3.2\sigma$
significance for $\bdsks$, with a branching fraction of
$\br(\bdsks)=(\brbdsks)\tim$.
No significant signal is observed in the $\bdskstar$ final state. 
The corresponding upper limit at the 90\% CL is
${\cal{B}}(\bdskstar) <\brbdskstar\tim$.
We also set the 90\% CL upper limits for the $V_{ub}$ suppressed
$\bdzbarkstar$ decays:
$\br(\bdbarkstar) <\brbdbarkstar\tim$ and
$\br(\bdsbarkstar) <\brbdsbarkstar\tim$.
The results obtained supersede 
our previous measurements~\cite{belle_br};
these results are consistent with,
and more precise than, the previous measurements.

We thank the KEKB group for the excellent operation of the
accelerator, the KEK Cryogenics group for the efficient
operation of the solenoid, and the KEK computer group and
the National Institute of Informatics for valuable computing
and Super-SINET network support. We acknowledge support from
the Ministry of Education, Culture, Sports, Science, and
Technology of Japan and the Japan Society for the Promotion
of Science; the Australian Research Council and the
Australian Department of Education, Science and Training;
the National Science Foundation of China under contract
No.~10175071; the Department of Science and Technology of
India; the BK21 program of the Ministry of Education of
Korea and the CHEP SRC program of the Korea Science and
Engineering Foundation; the Polish State Committee for
Scientific Research under contract No.~2P03B 01324; the
Ministry of Science and Technology of the Russian
Federation; the Ministry of Education, Science and Sport of
the Republic of Slovenia;  the Swiss National Science Foundation; 
the National Science Council and
the Ministry of Education of Taiwan; and the U.S.\
Department of Energy.

\end{document}

%% file: author-conf2004.tex
%%% Paper:    
%%% Journal:  summer 2004 conference papers (PRL format)
%%% Contacts: 
%%% Last revised on July 14, 2004 16:40:00 EDT
%%% Non-responding authors or those who said NO are commented out.
%%% ====================================================================
%%% Click the RELOAD button on your web browser to see the updated file.
%%% ====================================================================
%%% Use \input{author} to insert this material into your latex file.
%%%%% Force institutions to appear in alphabetical order when typeset.
\affiliation{Aomori University, Aomori}
\affiliation{Budker Institute of Nuclear Physics, Novosibirsk}
\affiliation{Chiba University, Chiba}
\affiliation{Chonnam National University, Kwangju}
\affiliation{Chuo University, Tokyo}
\affiliation{University of Cincinnati, Cincinnati, Ohio 45221}
\affiliation{University of Frankfurt, Frankfurt}
\affiliation{Gyeongsang National University, Chinju}
\affiliation{University of Hawaii, Honolulu, Hawaii 96822}
\affiliation{High Energy Accelerator Research Organization (KEK), Tsukuba}
\affiliation{Hiroshima Institute of Technology, Hiroshima}
\affiliation{Institute of High Energy Physics, Chinese Academy of Sciences, Beijing}
\affiliation{Institute of High Energy Physics, Vienna}
\affiliation{Institute for Theoretical and Experimental Physics, Moscow}
\affiliation{J. Stefan Institute, Ljubljana}
\affiliation{Kanagawa University, Yokohama}
\affiliation{Korea University, Seoul}
\affiliation{Kyoto University, Kyoto}
\affiliation{Kyungpook National University, Taegu}
\affiliation{Swiss Federal Institute of Technology of Lausanne, EPFL, Lausanne}
\affiliation{University of Ljubljana, Ljubljana}
\affiliation{University of Maribor, Maribor}
\affiliation{University of Melbourne, Victoria}
\affiliation{Nagoya University, Nagoya}
\affiliation{Nara Women's University, Nara}
\affiliation{National Central University, Chung-li}
\affiliation{National Kaohsiung Normal University, Kaohsiung}
\affiliation{National United University, Miao Li}
\affiliation{Department of Physics, National Taiwan University, Taipei}
\affiliation{H. Niewodniczanski Institute of Nuclear Physics, Krakow}
\affiliation{Nihon Dental College, Niigata}
\affiliation{Niigata University, Niigata}
\affiliation{Osaka City University, Osaka}
\affiliation{Osaka University, Osaka}
\affiliation{Panjab University, Chandigarh}
\affiliation{Peking University, Beijing}
\affiliation{Princeton University, Princeton, New Jersey 08545}
\affiliation{RIKEN BNL Research Center, Upton, New York 11973}
\affiliation{Saga University, Saga}
\affiliation{University of Science and Technology of China, Hefei}
\affiliation{Seoul National University, Seoul}
\affiliation{Sungkyunkwan University, Suwon}
\affiliation{University of Sydney, Sydney NSW}
\affiliation{Tata Institute of Fundamental Research, Bombay}
\affiliation{Toho University, Funabashi}
\affiliation{Tohoku Gakuin University, Tagajo}
\affiliation{Tohoku University, Sendai}
\affiliation{Department of Physics, University of Tokyo, Tokyo}
\affiliation{Tokyo Institute of Technology, Tokyo}
\affiliation{Tokyo Metropolitan University, Tokyo}
\affiliation{Tokyo University of Agriculture and Technology, Tokyo}
\affiliation{Toyama National College of Maritime Technology, Toyama}
\affiliation{University of Tsukuba, Tsukuba}
\affiliation{Utkal University, Bhubaneswer}
\affiliation{Virginia Polytechnic Institute and State University, Blacksburg, Virginia 24061}
\affiliation{Yonsei University, Seoul}
  \author{K.~Abe}\affiliation{High Energy Accelerator Research Organization (KEK), Tsukuba} % KEK
  \author{K.~Abe}\affiliation{Tohoku Gakuin University, Tagajo} % TohokuGakuin
  \author{N.~Abe}\affiliation{Tokyo Institute of Technology, Tokyo} % TIT
  \author{I.~Adachi}\affiliation{High Energy Accelerator Research Organization (KEK), Tsukuba} % KEK
  \author{H.~Aihara}\affiliation{Department of Physics, University of Tokyo, Tokyo} % Tokyo
  \author{M.~Akatsu}\affiliation{Nagoya University, Nagoya} % Nagoya
  \author{Y.~Asano}\affiliation{University of Tsukuba, Tsukuba} % Tsukuba
  \author{T.~Aso}\affiliation{Toyama National College of Maritime Technology, Toyama} % Toyama
  \author{V.~Aulchenko}\affiliation{Budker Institute of Nuclear Physics, Novosibirsk} % BINP
  \author{T.~Aushev}\affiliation{Institute for Theoretical and Experimental Physics, Moscow} % ITEP
% ...
% Lausanne
\collaboration{The Belle Collaboration}

%% file: belle-conf-0446.bbl
\begin{thebibliography}{99}
\bibitem{conj}{
The inclusion of charge conjugate states is implicit 
throughout this report. Note that we do not distinguish between final
states with $\ks$ and $K^0$.}
\end{thebibliography}
